\documentclass[12pt]{article}
\usepackage{latexsym}

\def\tr{{\rm tr}}

\def\ket#1{\mid~\!\!\!{#1}~\!\!\rangle}
\def\bra#1{\langle~\!\!{#1}~\!\!\!\mid}
\def\+-{\buildrel + \over -}

\def\QM{quantum mechanics }
\def\qm{quantum mechanics}
\def\Q{quantum }

\def\QMl{quantum-mechanical }
\def\qml{quantum-mechanical}

\def\${\enskip$}
\def\WF{wavefunction }
\def\wf{wavefunction}
\def\M{measurement }
\def\m{measurement}
\def\E{ensemble }
\def\e{ensemble}
\def\IS{individual-system }
\def\is{individual-system}

\begin{document}

{\large \bf \noindent Fleeting Critical Review of the Recent\\ Ontic Breakthrough in Quantum Mechanics}\\

\begin{quote}
{\noindent\small\rm Fedor Herbut}\\
{\noindent\scriptsize\it Serbian Academy
of Sciences and Arts, Belgrade, Serbia}\\

{\footnotesize \noindent The ontic breakthrough in \Q foundations, consisting of three theorems, that of Pusey, Barrett, and Rudolph (PBR), the Colbeck-Renner one, and Hardy's one, is shortly presented, together with various reactions. Some of the ideas involved are explained and/or commented upon. Thus, the \WF is proved real in three independent ways. Each of the theorems rests on more or less plausible assumptions, but they require more in-depth analyses.}\\

\end{quote}

 \normalsize \rm
\section{INTRODUCTION}

CONTENTS:

\noindent
2   On the seminal work of Harrigan and       Spekkens

\noindent
2.1 On the ontic model

\noindent
2.2 On the definition of ontic and
epistemic \Q theories

\noindent
2.3 More remarks on \Q reality

\noindent
3   On the PBR theorem and on reactions to it

\noindent
3.1 On the PBR theorem

\noindent
3.2 Leifer

\noindent
3.3 Drezet

\noindent
3.4 Schlosshauer and Fine

\noindent
3.5 On other reactions

\noindent
4   Colbeck and Renner

\noindent
4.1 Colbeck and Renner on the
possibility of improving \QM

\noindent
4.2 On reactions to the first,
no-improvement,  Colbeck-Renner
theorem

\noindent
4.3 The Colbeck-Renner theorem
claiming ontic nature of the \WF

\noindent
5   Hardy
's theorem claiming ontic
nature of the \WF

\noindent
6   Concluding remarks

\noindent
REFs

Let me start with an incisive quote of Jaynes  \cite{Jaynes}.

\begin{quote}
{\bf Ja:} "But our present (quantum mechanical) formalism is not purely
epistemological; it is a peculiar mixture describing in part
realities of Nature, in part incomplete human information about
Nature — all scrambled up by Heisenberg and Bohr into an omelette that nobody has seen how to unscramble. Yet we think
that the unscrambling is a prerequisite for any further advance
in basic physical theory. For, if we cannot separate the subjective
and objective aspects of the formalism, we cannot know what we
are talking about; it is just that simple."
\end{quote}

To put the approach adapted in this review metaphorically, I'll skate rather lightly over the surfaces of very deep pools because the 'pools' are so 'deep', that wherever I would dive, it would take a whole article to surface again. The reader will judge if my quotes are well chosen,and if my explanations and remarks are in order.

If the reader still feels puzzled with the meaning of my expression "fleeting critical review" in the title, he (or she) should have a look at the very recent thorough critical review of Leifer \cite{Leifer1} of 116 pages (an e-book) on the same subject to see a contrast.

\section{On the seminal work of Harrigan and Spekkens}

The \WF is actually an \E entity: it consists of the entirety of all possible probabilities \cite{Gleason}. If one wants to read the \IS information in it, one may find that there is nothing to read. (For a different point of view see Hetzroni and Rohrlich at the end of subsection 3.4 below.)\\

Harrigan and Spekkens (HS by acronym) have given \cite{Spekkens1} a precise {\bf ontic model} for \is , so-called {\bf ontic states} denoted by \$\lambda\$. To any given \WF \$\ket{\psi}\$ corresponds a set \$\Lambda_{\psi}\$ of {\bf all} sub\Q states \$\lambda\$ that can be the \IS byproduct of any
 preparation of the system in the state $\ket{\psi}\$. The distinct \$\lambda\$ may be thought to distinguish the individual systems when \$\ket{\psi}\$ appears \e wise.

The ontic model of HS seems to be an attempt  {\bf to begin to "unscramble" subjective from objective} (cf quote "Ja" in the Introduction). Finally, one can begin to try to define what is \Q reality.\\

\subsection{On the ontic model}

Let me quote their precise words.

\begin{quote}
{\bf Ha-Sp-1:} "{\bf Definition 1}. An ontological model of operational quantum theory posits an ontic state space \$\Lambda\$  and prescribes a probability distribution over  \$\Lambda\$ for every preparation
procedure P , denoted \$p(\lambda |P)\$, and a probability distribution over the different outcomes \$k\$ of a measurement \$M\$ for every ontic state \$\lambda\in\Lambda\$, denoted \$p(k|\lambda ,M)\$. Finally,
for all P and M, it must satisfy,

$$\int_{\Lambda } d\lambda p(k|M,\lambda )p(\lambda |P)=tr(\rho E_k),\eqno{(Ha-Sp-1)}$$

where \$\rho\$ is the density operator associated with P and \$E_k\$
is the POVM element associated with outcome k of M ."
\end{quote}

{\bf My comments}: The authors treat the most general \Q case with an arbitrary \Q state \$\rho\$. One is usually confined to pure states \$\rho\equiv\ket{\psi}\bra{\psi}\$, and to ordinary obserbales or PV (projector-valued) measures defining ordinary observables (special case of POV - positive-operator-valued - measures defining generalized observables).

Most authors who used the HS ontic model restricted their attention to \wf s. A preparation \$P\$ of a pure state \$\ket{\psi}\$ is better denoted by \$P_{\psi}\$. A given pure state can be prepared in many ways. Denoting the set of all ontic states corresponding to \$\ket{\psi}\$ by \$\Lambda_{\psi}\$, one obviously has \$\Lambda_{\psi}\equiv\cup_{all P_{\psi}}\Lambda_{P_{\psi}}\$, where "$\cup$" stands for the set-theoretical union.\\

The {\bf simplest} HS ontic model is the one in which each set \$\Lambda_{\psi}\$ contains only one element, and this element wold then necessarily be the state vector \$\ket{\psi}\$.

Formally, this follows from relation (Ha-Sp-1): \$p(\lambda|P_{\psi})\$ would have some Dirac function form (on the set of all \$\lambda\$ vales) implying
$$ p(k|M,\lambda )=\bra{\psi}Q(k|M)\ket{\psi},\eqno{(Ha-Sp-1)}'$$
where \$Q(k|M)\$ denotes the corresponding eigen-projector, and one has the standard \QMl expectation formula on the RHS (special case of RHS(Ha-Sp-1)). Taking into account all observables \$M\$ and all their eigenvalues \$k\$, we know from Gleason's theorem \cite{Gleason} that \$\lambda\$ on the LHS then defines \$\ket{\psi}\$ on the RHS.

The other extreme case, the dream of most hidden-variables seekers, is when to each state \$\ket{\psi}\$ do correspond different ontic states \$\lambda\$, but each of them 'knows all the answers', i. e., for each \$\ket{\psi}\$, for each \$\lambda\in\Lambda_{\psi}\$, and for each observable \$M\$, one has
\$p(k|M,\lambda )=\delta_{k,k_0}\$.

This wold reduce the randomness in measurement results  to ignorance, just like in classical physics. This time the ignorance would apply to the ontic state: we do not know which ontic state \$\lambda\$ will be the byproduct of a preparation \$P_{\psi}\$ of \$\ket{\psi}\$. But on, what I would call the {\bf sub\Q level}, which is inaccessible to \Q laboratory operations and beyond the \Q formalism, the measurement results are causal (deterministic).\\

It is remarkable that HS allowed for randomness in the ontic states (the probabilities \$p(k|M,\lambda )\$ in relation (Ha-Sp-1)). They have in this way separated the problem of \Q reality from that of the desire to reduce \Q randomness to ignorance as it is the case in classical physics.\\

\subsection{On the definition of ontic and epistemic \Q theories}

Later on in their article \cite{Spekkens1} the authors say:

\begin{quote}
{\bf Ha-Sp-2:} "{\bf Definition 4}. An ontological model is \$\psi$-ontic if for any pair of preparation procedures, \$P_{\psi}\$ and \$P_{\phi}\$, associated with distinct quantum states \$\psi\$
and \$\phi\$,we have"

$$p(\lambda |P_{\psi})p(\lambda |P_{\phi})=0\quad\mbox{for all}\quad \lambda .\eqno{(Ha-Sp-2)}$$
\end{quote}
\vspace{3mm}

Their crucial definition 4 states that a \Q model, which can also be an interpretation of \qm , is \$\psi-ontic\$, or equivalently, each
\Q state \$\rho\$ or, in particular, \$\ket{\psi}\$ in it is {\bf real}, if the sub\Q spaces \$\Lambda_{\psi}\$ and \$\Lambda_{\phi}\$ {\bf never}, i. e., for no value of \$\lambda\$, {\bf overlap}.

Then the ontic state \$\lambda\$ uniquely implies the \WF \$\ket{\psi}\$.

This definition makes perfect sense to our classically trained minds. Namely, in classical statistical physics we have overlapping probability distributions (e. g. Gaussians) over the phase space of the material point. They express our statistical knowledge about the system and are not a property of the individual system. These distributions are epistemic.

Contrariwise, if there is no overlap, like e. g. if the distribution is given in terms of the kinetic energy of a free material point, then the counterpart of \$\Lambda\$ is the set of all phase space points having the same absolute value of the linear momentum. These classical counterparts do not overlap, and the kinetic energy is a property of the individual mass point. This distribution is ontic. (See outlines of the criticism of Drezet in subsection 3.2 .)\\

Returning to the HS ontic model and to their definition of \$\psi$-ontic and \$\psi$-epistemic, let me state in precise terms how I understand their idea of ontic \wf s. The \WF is always epistemic in a general sense because it represents our knowledge about a \Q system. But if one says that it is HS ontic ($\psi$ -ontic), then one means that all physical information contained in the \WF applies to each individual system that is described by the wave function. Otherwise, if the \WF is HS epistemic ($\psi$-epistemic), then it contains only some information on the individual system (but it describes precisely and completely the ensemble, as it always does).

One should note that the ontic state \$\lambda\$ is not necessarily an entity that has no connection with the given \Q state \$\ket{\psi}\$. The de Broglie-Bohm theory may be a good example for this. The \WF is ontic, i. e., it describes the state of the \IS in it to such a high degree that the time evolution of the hidden variable, the position \$x\$ of the particle, depends on it. Here it appears that  \$\lambda =\ket{\psi}+x\$ (but see Drezet in 3.2 below for a different opinion).

One should keep in mind that de Broglie-Bohm theory is, perhaps, the most successful hidden-variable theory - cf \cite{Bohm}, \cite{Holland}, \cite{BHK}, but see also the critical article of Zeh \cite{Zeh}.
Hence, de Broglie-Bohm theory  is the first that comes to mind when discussing the possibility of a sub\Q reality.\\

Spekkens, who seems partial to an epistemic view of the \wf , writes (in his famous 'toy theory' \cite{Spekkens2}, (on p. 2, calling the ontic view "that in terms of hidden variables"):

\begin{quote}
{\bf Sp:} "This prompts the obvious question: if a quantum state is a state of knowledge, and it is not knowledge of hidden variables, then what is it knowledge about? We
do not at present have a good answer to this question. We
shall therefore remain completely agnostic about the nature of the reality to which the knowledge represented by quantum states pertains."
\end{quote}

It is clear that Spekkens does not doubt the existence of \Q reality, which is embodied in  \$\cup_{all \psi}\Lambda_{\psi}\$. He just favors no concrete idea what the ontic states \$\lambda\in\Lambda{\psi }\$ might be.

Spekkens' preference of the epistemic view is much based on work of Fuchs and coauthors (see \cite{RMPreview} and the references therein).

But it is also a relevant fact
that HS mentioned Einstein in the title of their article \cite{Spekkens1}. Einstein was known to view the \WF only \e wise, and he expected that additional information should define the \IS states in some way. One can say that HS gave a concrete form to Einstein's expectation, and thus {\bf opened the door} for new and relevant exact research.\\

\subsection{Remarks on \Q reality}

Before we proceed, we raise the question of {\bf \Q reality}. If \QM is {\bf \$\psi$-ontic}, then the answer is that the carrier of \Q reality  is the {\bf \wf }: each \IS state is always an ontic state \$\lambda\$, and it implies a unique \Q state \$\ket{\psi}\$. In other words, the \IS state has a definite \WF (just like when the \$\lambda\$ are superfluous - cf the first extreme case at the end of the previous subsection).\\

If, on the other hand, \QM is {\bf \$\psi$-epistemic}, then we are in trouble again. The \WF contains information on the sub\Q reality in an ill defined way. And \Q reality rests on this information and on the entirety of the sub\Q reality in some way.

Let me try to explain this in terms of Plato's caves. Let me compare the \wf s to the shadows on the wall of the cave and \$\lambda\$ to the animal running outside the cave and making the shadow. Then in \$\psi$-ontic \qm , though one and the same shadow appears for any from a bunch of animals, we cannot tell apart the individual ones, but different shadows are certainly cast by animals from a different bunch. So our shadows carry reality, though in an unsharp way.

If \QM is \$\psi$-epistemic, then we have a complete chaos in the relation between shadow and animal. One and the same animal can cast different shadows, and the same shadow can, like in the former case, be due to different animals. A nightmare for the caveman. He had best turn his back on the wall; and that is what the non-realists do.

The bottom line is that if \QM would  turn out \$\psi$-epistemic in the end, then objective and subjective would remain scrambled (cf quote "Ja" in the Introduction), and this might cause the specter of anti-realism, which has been lurking all the time, rise again.\\

As I have pointed out, HS, by allowing for a possible statistical nature of the sub-quantum predictions \$p(k|M,\lambda )\$ of \M results, "unscrambled" at least the so-called \Q \M paradox from that of \Q and possible sub\Q reality.

Let me elaborate how the \M paradox looks like in the HS ontic model in case of \$\psi$-ontic and \$\psi$-epistemic \qm . The \M of a given observable \$M\$ is done ensemblewise. But each ensemble consists of individual systems. Each if these is in an ontic state \$\lambda\$ that is determined only with a probability \$p(\lambda )\$. This is the first randomness. Perhaps one might say it is the \Q one: one does not know, when preparing \$\ket{\psi}\$, in which ontic state \$\lambda\$ the individual system will land.

Next, the ontic state may not determine a definite result \$k\$; there may be positive probabilities \$p(k|M,\lambda )>0\$ for several possible results (cf relation (Ha-SP-1) above). This is the second, the sub\Q randomness.

Thus we have the \QMl \M paradox replaced by two randomnesses, both completely unknown theoretically and experimentally.

These few remarks should suffice for bringing home the idea that the ontic breakthrough, which I am going to report on (and comment on) in the rest of this article, raises great hopes for a \$\psi$-ontic \qm , and herewith for a clear and simple idea at least of \Q reality.\\

\section{On the PBR theorem and on reactions to it}

The recent important work of Pusey, Barrett, and Rudolph  \cite{Pusey},
referred to as of PBR, utilized the expounded sub\Q framework of Harrigan and Spekkens to present
a specific theorem in which they give a proof for the ontic interpretation of the \WF in terms of the \QMl formalism.

There were numerous reactions to it.\\

\subsection{On the PBR theorem}

We can read in the the first column of the first page of their article:

\begin{quote}
{\bf Pu-Ba-Ru:} "This Article presents a no-go theorem: if the quantum state merely represents information about the real physical state, then experimental predictions are obtained that completely
contradict those of quantum theory.

The argument depends on few
assumptions. One is that a system has a 'real physical state' — not necessarily completely described by quantum theory, but objective and independent of the observer. This assumption only needs
to hold for systems that are isolated, and not entangled with other systems.

Nonetheless, this assumption, or some part of it, would be denied by instrumentalist approaches to quantum
theory, wherein the quantum state is merely a calculational tool for making predictions concerning macroscopic measurement outcomes.

The other main assumption is that systems that are prepared independently have independent physical states."
\end{quote}

As far as I understand, the authors mean by "physical state" a subquantum \IS state, a concrete 'value' of the entity \$\lambda\$ in the HS ontic model.\\

It is remarkable that PBR allow for the possibility that there is no subquantum reality, i. e., that \$\Lambda\$ is a set of one element: \$\Lambda\equiv\{\lambda\equiv\ket{\psi}\}\$.\\

I will say no more about their work here. I'll rather turn to reactions to their discovery.\\

\subsection{Leifer}
The PBR article immediately, even in its arxiv form, gave rise to violent reaction by Leifer \cite{Leifer2}. Let me quote from Leifer's detailed discussion.

\begin{quote}
{\bf L1:} ""I think that the PBR result is the most significant constraint on hidden variable theories that has been proved to date. It provides a simple proof of many other known theorems, and it supercharges the EPR argument, converting it into a rigorous proof of nonlocality that has the same status as Bell's theorem."
\end{quote}

Later on he writes:
\begin{quote}
{\bf L2:} I am not sure I would go as far as Antony Valentini, who is quoted in the Nature article saying that it is the most important result since Bell's theorem, or David Wallace, who says that it is the most significant result he has seen in his career. Of course, these two are likely to be very happy about the result, since they already subscribe to interpretations of quantum theory in which the quantum state is ontic (de Broglie-Bohm theory and many-worlds respectively) and perhaps they believe that it poses more of a dilemma for epistemicists like myself than it actually does."\\
\end{quote}

\subsection{Drezet}

Though Leifer has conveyed the impression that an adherent to the de Broglie-Bohm interpretation of \qm , like Valentini, can be expected only to applaud to the BPR theorem, a kind of surprise came from another adherent  Drezet \cite{Drezet1}, \cite{Drezet2}. Let me quote from his latter article.

In the very Abstract, having in mind the question "Can quantum mechanics be considered as statistical?" in the title of the article, the author says:

\begin{quote}
{\bf Dr-1:} "The answer to this question is 'yes it can!'" further "contrarily to the PBR claim the epistemic approach is in general not disproved by their 'no-go' theorem."
\end{quote}

Clearly,  by the term 'statistical' in the title the author means 'at least partially epistemic'. In the passage before the Conclusion he says:

\begin{quote}
{\bf Dr-2:} "The theorem proposed by PBR is thus simply not general enough. It fits well with the XIXth-century-like hidden variable
models but it is not in agreement with a neo-classical model such as the one proposed by de Broglie and Bohm."
\end{quote}

Finally, in the Conclusion of the article we can read:

\begin{quote}
{\bf Dr-3:} "... PBR ... ruins the old-fashion hidden-variable approach in a nice way and show that there are some fundamental differences between classical XIXth century physics and the neo-classical mechanics proposed by Bohm and others. Both are based on realism. Both are admitting an ontic and epistemic parts. But now the wave function is part of the dynamic all the way long since it gives a contribution to the ontic state
which subsequently affects the dynamic of the \$\lambda$-particle."
\end{quote}

Drezet compares the sub\Q space, i. e., the union  \$U_{all\psi }\Lambda_{\psi }\$, with the classical space of the Liouville theorem. This might contain the seeds of a deeper understanding of the limitations of the PBR theorem.

Incidentally, the very fact that the PBR theorem is formulated in the HS ontic model (cf subsection 2.1), which is very plausible to our classically trained minds, should make us suspicious in view of our past failures with classical intuition in the \Q world. We are now entering a hypothetical sub\Q reality, and shouldn't we be cautious not to perpetrate a misuse of our classical intuition because we know no better? Thus, Drezet's critical approach is promising. But, perhaps due to the superficiality of my reading, I do not have the impression that he has proved his point in \cite{Drezet2}.\\

I may be completely wrong, but I think I see one problem that now may face the adherents to de Broglie-Bohm theory. It is a fact that if the \WF is ontic, as claimed in the PBR theorem, then every piece of physical information contained in the \WF is valid for the individual system (cf end of 2.1). Since the \WF actually consists of probability distributions, the latter must have an \IS meaning like, e. g., in my study \cite{FHStud}, where I have extended the delocalization concept from position to the spectra of all \QMl observables under the term 'degree of presence'.

It is my understanding that the adherents to de Broglie-Bohm theory have a different idea because they believe in the sub\Q point-like localization of every particle, and then delocalization does not make sense unless it is statistical, i. e., unless it appears only ensemblewise. Thus, the \WF should be ontic for them, but not completely ontic. Such a 'loosened up' onticity is, perhaps what they need, and they may hope to find it in a sub\Q frame that is broader than that of Harrigan and Spekkens in the sense that it should allow partly ontic and partly epistemic \wf s.

Whether my conjecture about Drezet's motivation for his reaction to the PBR theorem is is right or not, his subsequent  endeavor \cite{Drezet3} seems to hit the target.

In \cite{Drezet3} the author claims, if I understand it correctly, that he has found a tacit dynamical assumption that is indispensable for the validity of the PBR theorem, and which the de Broglie-Bohm theory does not satisfy. So he seems to have proved that the \WF is not completely ontological in the latter theory.

In his own words, this reads as follows
(see his "Conclusion"):

\begin{quote}
{\bf Dr-4:} "To conclude, we analyzed the PBR theorem and showed that beside the important independence criterion ...
there is a second fundamental postulate associated with \$\psi$-independence at the dynamic level ... . We showed that by abandoning this prerequisite the PBR conclusion collapses. ..."
\end{quote}

So, Drezet seems to have succeeded to save de Broglie-Bohm theory from the claws of the PBR no-go theorem because, as far as I understand, the former does not satisfy this second (dynamical) {\bf tacit} assumption of the theorem. Thus, the \WF can be partially ontic and partially epistemic in this hidden-variable theory.

If Drezet's proof is correct, it may be of great significance. Apparently, it may give a profound view both of the PBR theorem and of de Broglie-Bohm theory.\\

\subsection{Schlosshauer and Fine}

Schlosshauer and Fine analyzed consequences of the PBR theorem on hidden-variable theories \cite{MAX1} and \cite{MAX2}. Their results are certainly of great importance, but to present any of them would require a detailed knowledge of hidden variables - a rich and elaborate domain of foundational quantum research.

In \cite{MAX2} they derived puzzling implications of the theorem.
In the very Abstract they  wrote:

\begin{quote}
{\bf Sch-Fi:} "Building on the Pusey-Barrett-Rudolph theorem, we derive a no-go theorem for a vast class of deterministic hidden-variables theories, including those consistent on their targeted domain. The
strength of this result throws doubt on seemingly natural assumptions (like the "preparation independence" of the Pusey-Barrett-Rudolph theorem) about how "real states" of subsystems compose
for joint systems in non-entangled states. This points to constraints in modeling tensor-product states, similar to constraints demonstrated for more complex states by the Bell and Bell–Kochen– Specker theorems."\\
\end{quote}

This sounds dramatical to me. To understand the meaning of their words, let us be reminded of the famous theorem of Bell \cite{BELL}, in which it has been proved that real local hidden variables contradict \qm . Numerous experimental investigations, which began with \cite{FC}, \cite {ASPECT} etc., have proved subsequently that it is the \QMl formalism, and not the hypothesis of local sub\Q reality, that is true in nature.

One can speculate that in the HS ontic model (cf subsection 2.1) the hypothesis of 'local subqantum reality' can be formulated as follows: Whatever the two-particle \WF \$\ket{\Psi}_{12}\$, each of the particles \$i=1,2\$ has a sub\Q \IS state \$\lambda_{\rho_i}\enskip\Big(\in\Lambda_{\rho_i}\Big)\$,
where \$\rho_i\equiv\tr_j\Big(\ket{\Psi}_{12}\bra{\Psi}_{12}
\Big)\$, \$i\not= j,\enskip i,j=1,2\$, are the subsystem state operators (reduced density operators) of the particles. Possibly, the preparation of \$\ket{\Psi}_{12}\$ may establish a correlation between \$\lambda_{\rho_1}\$ and \$\lambda_{\rho_2}\$.

A weaker form would require the above only if the particles in the given bipartite state are sufficiently  apart so that they do not interact. An even more weakened hypothesis would posit the above claim only if the particles have a spacelike distance in the Minkowski space, when they cannot interact in any way.

Now we know that none of these can be upheld. There is non-locality in the sub\Q reality (if there is a sub\Q reality). In terms of the present notions it would mean that the two-particle \WF has one common \IS sub\Q state \$\lambda_{\ket{\Psi}_{12}}\$ as long as the \WF is entangled (correlated \qml ly).

Now we can grasp the impact of the hypothetically put conjecture of Schlosshauer and Fine
(see the above quote Sch-Fi). It says, if I understand it correctly, that it can happen that, even if the two-particle \WF is uncorrelated \$\ket{\Psi}_{12}=\ket{\psi}_1\otimes\ket{\phi}_2\$,
one can have only a common (nonlocal) sub\Q \IS state \$\lambda_{\ket{\Psi}_{12}}\$. Possibly, N particles in an uncorrelated \Q state may have only a common ontic state \$\lambda_{1\dots N}\$ etc. Namely, this is to what amounts non-validity of the PBR initial assumption that independently prepared particle states have independent sub\Q states.

Such colossal sub\Q non-locality would seem amazing because in Bell's case, when there is \Q entanglement in the bipartite \Q \wf , and it is, as a rule, due to some interaction in the past, we can somehow blame this interaction for the creation of \$\lambda_{\ket{\Psi}_{12}}\$ instead of
\$\lambda{\rho_1}\$ and \$\lambda{\rho_2}\$. But, at first glance, there would be no dynamical mechanism for \$\lambda_{\ket{\Psi}_{12}}\$ if lack of correlation on the \Q level could be accompanied by strong correlation on the sub\Q level.

On the other hand, if we face up to the fact that we actually know very little about sub\Q reality (we do not know if it exists at all), then we must consider the following argument. All particles have interacted in some way in the past, and that may have established strong sub\Q correlations (like those suggested by Bell's theorem). If one makes two independent preparations of \$\ket{\psi}\$ and \$\ket{\phi}\$, one destroys the \Q correlations of each of the particles with its entire environment (a pure state cannot be entangled with its environment). But we do not really know if this implies also the destruction of the sub\Q correlations that existed prior to the preparations.\\

\subsection{Other Reactions}

In a quick reaction Hall \cite{Hall} made an apparently serious  attempt to weaken the PBR assumptions, on which their theorem rests, to generalize it.\\

The PBR theorem gave rise to proposing experimental investigation by Miller \cite{Miller}: "... we propose alternative experimental protocols which lead to the PBR result for a special case and a weaker PBR-like result generally. Alternative experimental
protocols support the assumption of measurement independence required for the PBR theorem" - as the author states in the abstract.\\

Wallden \cite{Wallden} was inspired to investigate whether the PBR theorem is valid for the interpretation pf \QM that he is interested in. He says (in the abstract): "Each interpretation of quantum theory assumes different ontology and one could ask if the PBR
argument carries over. Here we examine this question for histories formulations in general ... While the PBR argument goes through up to a point, the failure
to meet some of the assumptions they made does not allow one to reach
their conclusion."\\

Aaronson et al.  apparently followed PBR and  tried to add to the blow to epistemic interpretations of \qm . They
had an interesting result \cite{Aaronson}. They say (in their abstract): "In this paper, we prove that even without the Kochen-Specker or PBR assumptions, there are no \$\psi$-epistemic theories in dimensions \$d\geq 3\$ that satisfy two reasonable conditions: (1) symmetry under unitary transformations, and (2) "maximum nontriviality" (meaning that the
probability distributions corresponding to any two non-orthogonal states overlap). This no-go
theorem holds if the ontic space is either the set of quantum states or the set of unitaries."

The authors do not know if their result can be extended to infinite-dimensional Hilbert spaces. Thus, they have only an indication for \qm , but an interesting one.\\

Another attempt to generalize the PBR theorem is made by Mansfield \cite{Mansfield}. He says in his abstract about his work: "... a new characterization of non-locality and contextuality is found. Secondly, a careful analysis of preparation independence, one of the key assumptions of the PBR theorem, leads to an analogy with Bell locality, and thence to a proposal to weaken it to an assumption of 'no-preparation-signalling' in analogy with no-signalling. This
amounts to introducing non-local correlations in the joint ontic state, which is, at least, consistent with the Bell and Kochen-Specker theorems. The question of whether the PBR result can be strengthened to hold under this
relaxed assumption is therefore posed."

The direction in which Mansfield moves comes close to the case that I have discussed at the end of the preceding subsection. I understand this is the beginning of an investigation. We can only wish the author much success.\\

Bandyopadhyay et al. have written \cite{Band} an article under the title "Conclusive exclusion of quantum states". They say that they were motivated for this work by the PBR theorem.\\

Emerson et al, wrote up an article \cite{Emerson} in which they claim to have delivered a death blow to the PBR theorem. They put it like this (in their abstract): "We show that PBR's assumption of independence encodes
an assumption of local causality, which is already known to conflict with the predictions of quantum
theory via Bell-type inequalities."

It seems to me that there is room for violent confrontation of arguments. It would be a bad thing if one just ignored this provocation (or enlightenment).\\

Fortunately, we can end this section by a support for the PBR theorem. Hetzroni and Rohrlich \cite{HR} are concerned with protective \m s. They write (in their abstract):

"... protective measurements demonstrate
that a quantum state describes a single system, not only an ensemble of systems, and reveal a rich ontology in the quantum state of a single system. We discuss in what sense protective measurements
anticipate the theorem of Pusey, Barrett, and Rudolph (PBR), stating that, if quantum predictions are correct, then two distinct quantum states cannot represent the same physical reality."

\section{Colbeck and Renner}

The second important event in \Q foundations came from Colbeck and Renner (CR by acronym). They published two surprising results.

\subsection{Colbeck and Renner on the possibility of improving \QM }

Colbeck and Renner published in Nature Communications in 2011 an article \cite{Colbeck1} under the provocative title "No extension of quantum theory can have improved predictive power".

To begin with, I was baffled with the title. The "predictive power of \Q theory" are the probability distributions that a \Q state predicts for the \M of all possible observables. How can one improve on that. As far as I know, if you are given two probability distributions, you can hardly tell which is better, which is an improvement on the other (unless they have finite entropies and the latter is smaller). The only clear improvement on a probability distribution is a Kronecker distribution, i. e., a certainty.

It turned out, reading the article, that the authors actually meant by "no improvement" "no change at all".

The CR theorem at issue
, as all theorems, has an input: some assumptions the validity of which implies the claim of the theorem. It struck me first that their assumptions were very broad: free will (on part of the experimenter to choose any experiment). My first thought was that this assumption is usually tacitly always present.

It turned out next that the "free-will assumption" is a true quagmire. Ghirardi and Romano criticized \cite{Ghir1} the way how CR defined "free will". Leegwater wrote a master thesis \cite{Leeg} on the article of CR at issue, and also he seemed to pay most attention to the question of "free will". Then I have consulted my archive, and I found a large number of interesting and apparently important articles on free will that preceded the CR article. Let me share only some of them \cite{AJP}, \cite{Conwey}, \cite{'t Hooft} .

I came to the conclusion that the article in question should be left alone for some time till some idea of free will will be widely accepted. But I have changed my mind when I Googled up Laudisa's comments \cite{Laudisa} (pp. 16-17).

Laudisa called my attention to the fact that CR envisaged, in their article at issue, the hypothetical improvement of \QM by adding a completely general new piece of information \$\Xi\$ to a quantum state, and showing that it was irrelevant; the predictive power was unchanged. But they required \$\Xi\$ to satisfy some innocuous conditions, among them that it was "\textbf{always accessible}" and \textbf{static}, i. e., not changing in time.

Returning to the HS ontic model (cf subsection 2.1), the sub\Q \IS state \$\lambda\$ may be hoped to be, at least partly, some day accessible and static. Let us denote this part by \$\lambda_1\$. Let the remaining part be \$\lambda_2\$. We have \$\lambda =\lambda_1+\lambda_2\$. What the article of CR seems to prove is that if some day \$\lambda_1\$ is added to the \Q state in the hope of obtaining a better theory, one will still have the same theory (with the useless \$\lambda_1\$ added). About \$\lambda_2\$ the theorem in question apparently says  nothing.\\

The well-known hidden-variable theory of de Broglie and Bohm is mentioned in the very article of CR, which is being discussed. It is commented that the former, as a result of the latter, apparently tacitly  violates free will. Laudisa points out that this violation need not be the case; it may be that the implication of the ontic CR theorem is that in de Broglie-Bohm theory the hidden-variable \$\lambda\$, which is the exact position of the particle, is not static, and it is, most likely, doomed to be never accessible (\$\lambda =\lambda_2\$ in our notation).\\

\section{On reactions to the first, no-improvement,  Colbeck-Renner theorem}

Vona and Liang \cite{Vona} utilized the first Colbeck-Renner theorem (on no improvement) to obtain some results on Bell's theorem. They say in their abstract: "... making use of a recent result by Colbeck and Renner
[Nat. Commun. 2, 411 (2011)], we then show that any nontrivial account of these correlations in
the form of an extension of quantum theory must violate parameter independence. Moreover, we
analyze the violation of outcome independence of quantum mechanics and show that it is also a
manifestation of nonlocality."

Forster \cite{Forster} saw an important implication of the first Colbeck-Renner theorem (on no improvement) to Bell's theorem. He says in his Introdiction: "Colbeck and Renner themselves do not relate their theorem directly to Bell's theorem at all, so here I present a version of the Colbeck-Renner theorem that makes the relationship clear and explicit.  I will refer to this version of the Colbeck-Renner theorem as the Stronger Theorem." In his abstract we find: "... the new theorem does not assume Outcome Independence".\\
    

\subsection{The Colbeck-Renner theorem claiming ontic nature of the \WF }

In  \cite{Colbeck2} (cf also the improved version \cite{Colbeck3}) the authors claim that they have proved that, under the general assumption of free choice of \m , which amounts to postulating free will (as it is generally done), they obtain the ontic nature of \wf s (like PBR do).

The key assumption is that of free will. The way it was used by the authors was criticized by Ghirardi and Romano \cite{Ghir2}, to which they wrote a short answer \cite{Colbeck4}, in which they say (in the abstact):

\begin{quote}
{\bf Co-Re:} "We argue that the concepts of "freedom of choice" and of "causal order" are intrinsically linked:
a choice is considered "free" if it is correlated only to variables in its causal future."
\end{quote}

It is hard to see to what extent the Colbeck-Renner ontic theorem has a "no-go' character, i. e., what kind of hidden variables are excluded by it. The authors compare \cite{Colbeck3}, on their first page, their assumptions with those of PBR and they say "Here we show that the same conclusion can be reached without imposing any internal structure on \$\Lambda\$. More precisely, we prove that \$\Psi\$ is a function of any complete set of variables that are compatible with a notion of "free choice"..."

But in the abstract of the same article they say "... "free choice". This notion requires that certain
experimental parameters, which according to quantum theory can be chosen independently of other
variables, retain this property in the presence of \$\Lambda\$."

This sentence is unclear to me. How can "free choice" not retain this property in the presence of \$\Lambda_{\ket{\psi}}\$? Does this imply some influence of \Q procedures on the sub\Q reality (as it is the case both in the PBR theorem and in that of Hardy - cf the next section)? I hope it does not.

Anyway, if the reader wants to clarify these difficult questions, he (or she) is advised to look it up in the detailed analysis of Laufer \cite{Laufer2}.

I expect that it will finally turn out that this second, ontic, CR theorem is widely acceptable. Then it may become no less important than the PBR theorem, which gave rise to so many reactions. I was not able to find any reaction to the ontic CR theorem except the mentioned Ghirardi and Romano one up to the thorough critical study of Leifer \cite{Leifer1}. This is a pity because it certainly deserves more attention.\\

\section{Hardy's theorem claiming ontic nature of the \WF }

A third interesting study claiming the ontic nature of the \WF was done by Hardy \cite{Hardy}.

Within the sub-\Q framework of Harrigan and Spekkens (cf 2.1), the author says (in the abstract):

\begin{quote}
{\bf Ha:} "The preparation of a quantum state corresponds to a distribution over the ontic states, \$\lambda\$. If we make three basic assumptions, we can show that the distributions over ontic states corresponding to distinct pure states are nonoverlapping. This means that we can deduce the quantum state from a knowledge of the ontic state. Hence, if these assumptions are correct, we can claim that the quantum state is a real thing (it is written into the underlying variables that describe reality). The key assumption we use in this proof is ontic indifference — that quantum transformations that do not affect a given pure quantum state can be implemented in such a way that they do not affect the ontic states in the support of that state."
\end{quote}

What strikes me, if "ontic indifference" is valid or not is hard to see. Nevertheless, Hardy's theorem is a valuable contribution to the ontic breakthrough.\\

\section{Concluding remarks }

The mentioned famous 'toy theory' of Spekkens \cite{Spekkens1} (a model for \QM of restricted resemblance with the latter) is now a rare example of an epistemic theory. (The sets of sub\Q states \$\Lambda_{\psi }\$ are overlapping in it).

Turning to exact theories, we seem to have definitely entered a phase of development in \Q foundations in which the ontic interpretations of \QM get the upper hand over the epistemic ones.

To my mind, the basic question is {\bf if there exists a sub\Q reality or not}. It is not a question if we want to introduce hidden variables or not. It is a question if they exist in nature, i. e., if all \$\Lambda_{\ket{\psi}}\$ as non-empty sets, which by definition comprise {\bf all} ontic \IS states associated with \$\ket{\psi}\$, contain more than one element \$\lambda =\ket{\psi}\$ (cf the "simplest HS ontic model" in subsection 2.1). If the "simplest" model were true in nature, then, if I understand the PBR theorem correctly, it would be proven that \QM is \$\psi$-ontic.

But how can we get an answer, when the hypothetical sub\Q reality defies experimental accessability so far?

What one does, one assumes that there is a sub\Q reality, and one tries to find out, utilizing the \Q formalism in many ingenious ways, how it cannot be. Hence the no-go theorems.

Both the PBR and Hardy's theorems are no-go theorems. They exclude classes of hidden variables and thus they bring us closer to an idea what they could be.

It seems to me that the second, ontic, Colbeck-Renner theorem carries the greatest push towards onticity in the ontic breakthrough in the second decade of this century. It seems to be of least "no-go" nature (cf 4.3).\\

The archive version of this review conists of two parts. The second \cite{ARX2} corresponds to this article. The first \cite{ARX1} has the following title:  "On Historical Background to the Ontic Breakthrough.I Polemic Defense of Quantum Reality"\\

The interested reader is advised to read Leifer's analytical study of the ontic breakthrough \cite{Leifer1}. The present fleeting encounter with the subject may just hopefully serve as an appetizer for more serious study.\\


\begin{thebibliography}{99}
\bibitem{Drezet2}
A. Drezet, "Can quantum mechanics be considered as statistical? an analysis of the PBR theorem" arxiv:1203.2475v2 .
\end{thebibliography}
\end{document}